# Satellite Quantum Communication
# via the Alphasat Laser Communication Terminal

*Quantum signals from 36 thousand kilometers above Earth*

Dominique Elser, Kevin Günthner, Imran Khan, Birgit Stiller, Christoph Marquardt, Gerd Leuchs
Quantum Information Processing Group (QIV)
Max Planck Institute for the Science of Light (MPL)
Erlangen, Germany
and University of Erlangen-Nuremberg
Dominique.Elser@mpl.mpg.de

Karen Saucke, Daniel Tröndle, Frank Heine, Stefan Seel, Peter Greulich, Herwig Zech
Tesat-Spacecom GmbH & Co. KG
Backnang, Germany

Björn Gütlich, Ines Richter, Rolf Meyer
Space Administration
German Aerospace Center (DLR)
Bonn, Germany

*Abstract*—By harnessing quantum effects, we nowadays can use encryption that is in principle proven to withstand any conceivable attack. These fascinating quantum features have been implemented in metropolitan quantum networks around the world. In order to interconnect such networks over long distances, optical satellite communication is the method of choice. Standard telecommunication components allow one to efficiently implement quantum communication by measuring field quadratures (continuous variables). This opens the possibility to adapt our Laser Communication Terminals (LCTs) to quantum key distribution (QKD). First satellite measurement campaigns are currently validating our approach.

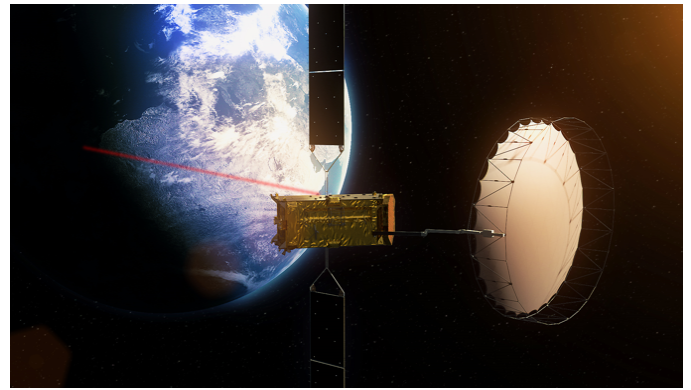

Figure 1: Alphasat I-XL is flying in geostationary Earth orbit (GEO) since 2013. The satellite carries a Laser Communication Terminal (LCT) for optical communication with satellites in low Earth orbit (LEO) and with ground stations. The LCT encodes information in a binary phase-shift keying (BPSK) alphabet using the two coherent quantum states $|\alpha\rangle$ and $|-\alpha\rangle$. Picture: ESA

## I. Introduction

The backbone of today's data communication is provided by optical fiber systems based on coherent communication [1]. Using the homodyne principle, known from radio communication, the signal is mixed with a local oscillator in order to filter out the background and detect small signals. In places where no optical fiber is installed, free space optical communication offers convenient alternative solutions. This applies especially to space optical communication, allowing for devices lighter and smaller than radio frequency links.

Tesat-Spacecom and the German Aerospace Center (DLR) have developed and successfully verified Laser Communication Terminals (LCTs) for space optical communication. LCTs of the first generation are flying on the spacecraft TerraSAR-X and NFIRE in low Earth orbit (LEO) since 2007. Coherent optical communication between these LCTs has been verified in 2008 [2] and LEO-to-ground communication in 2009 [3] . The second generation of LCTs is flying on Alphasat in geostationary Earth orbit (GEO) since 2013 (see Fig. 1) and on Sentinel-1a and -2a since 2014 and 2015, respectively. GEO-LEO optical coherent communication has been verified in 2014 [4] and GEO-to-ground communication is currently being verified [5].

The Max Planck Institute for the Science of Light (MPL) in collaboration with the University of Erlangen-Nuremberg has implemented coherent (continuous-variable) quantum communication systems since 2000 [6]. In 2008, MPL has verified the feasibility of continuous-variable quantum

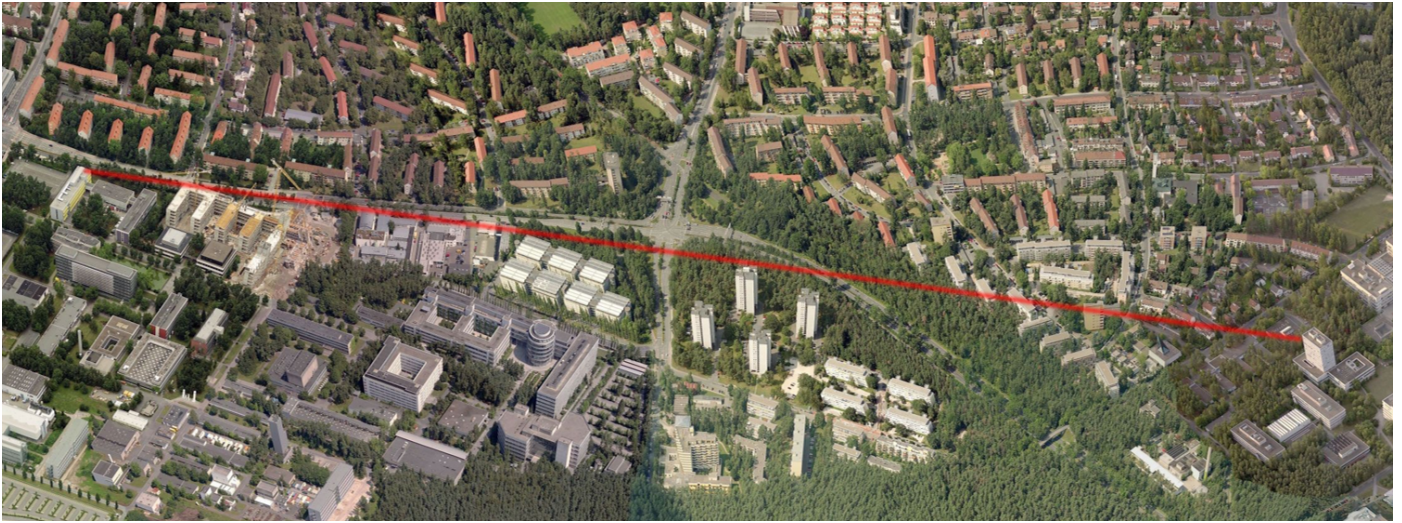

Figure 2: Free space quantum communication across Erlangen, Germany, serves as a testbed for coherent satellite quantum communication. The sender is located on the roof of the Max Planck Institute for the Science of Light (left). The receiver is within a building of the University of Erlangen-Nuremberg (right). Picture: Bing Maps

communication through turbulent atmosphere [7]. A continuous-variable free space quantum link (see Fig. 2) is operating in Erlangen since 2013 [8, 9].

The MPL quantum communication system and the Tesat/DLR LCTs are based on the same principle of coherent optical communication. This opened up a unique shortcut to implement space borne quantum communication at a cost much lower than other ambitious programs around the world [10].

## II. Principles of Coherent Quantum Communication

We can define a mode of the electromagnetic field by its frequency, wave vector, spatial and temporal shape and its polarization. In order to convey information via this mode, it has to carry some excitation. Depending on the excitation, the mode can be in different quantum states. Light from stars including the Sun is characterized by a thermal state, ruled by black body statistics. Coherent laser light produces in the ideal case coherent states, as introduced by Glauber [11]. Furthermore, single photon states or Fock states have found applications in quantum optics.

Since modern telecommunication technology uses coherent detection of coherent states, which are pure quantum states of light, we pursue this way for implementing satellite quantum communication. We can represent coherent states by their uncertainty area in a phase space diagram (see Fig. 3). The distance of the state from the origin corresponds to its amplitude and the angle to its phase. If we want to use coherent states to transmit information, we chose their distance in phase space such that the overlap of their uncertainty areas is small and the states are well distinguishable. For quantum key distribution (QKD), on the other hand, we want to exploit the fact that an eavesdropper cannot distinguish unambiguously between two non-orthogonal coherent states in phase space. In order to achieve non-orthogonal quantum states we move the signal states close to one another, e.g. at the origin in phase space, such that the quantum noise is of the order of the signal amplitude. Using this principle, we can implement satellite quantum key distribution (QKD) in the following way:

1. The sender on the satellite generates a sequence of true random numbers by using a quantum random number generator (QRNG). We can implement a QRNG by a homodyne measurement of an unmodulated signal (vacuum state) [12].

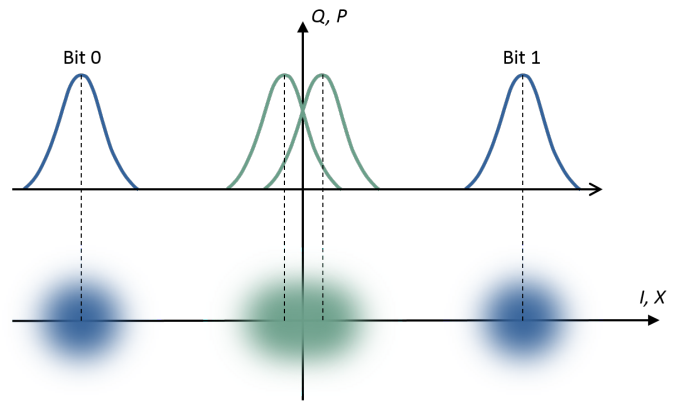

Figure 3: Phase space diagram of binary phase shift keying (BPSK) for data communication (blue) and for quantum communication (green) displaying all the states discussed below. The two axes of the diagram correspond to the in-phase component $I$ and the in-quadrature component $Q$, also known as $X$ and $P$ quadratures. The two states have the same distance from the origin, hence the same amplitude and a phase difference of $\pi$. Due to Heisenberg's uncertainty principle the states must occupy a minimum area in phase space. The projection of the in-phase component shows the overlap of the uncertainty areas. In data communication (blue), the overlap has to be small in order to distinguish between bits. In quantum communication (green), the overlap is harnessed to make eavesdropping impossible. A homodyne measurement yields a random value out of a continuous set within the uncertainty areas. Therefore, this method is known as continuous-variable quantum key distribution (CV-QKD), in contrast to discrete variables (clicks of single photon detectors).

2. The sender uses these random numbers to generate a random modulation pattern consisting of weak coherent states (green areas in Fig. 3).
3. The receiver performs homodyne detection of the quantum signal and analyses the measurement results with statistical means.
4. From the analysis of the quantum signal, the receiver deduces the amount of data post-processing (error correction and privacy amplification) to distill a secure key out of the raw data. The computing power for this post-processing can be concentrated at the ground station [13].
5. The satellite repeats this process with each user at different ground stations such that the satellite holds a shared secret key with each of these users. Any two users on ground can then request the satellite to broadcast the XOR of their keys. From this broadcast, only the two legitimate users can deduce a shared secret key.
6. The users encrypt their data communication with their key.

Keys are produced under good weather conditions and stored until used. Therefore, no real-time availability is required and weather has only a slight influence on the performance of the network. For the transmission of encrypted data, any channel with high availability can be used.

## III. BUSINESS CASE: LONG-HAUL QUANTUM KEY DISTRIBUTION

A transformation of the secure communication market is imminent in the near future. It is well known that the currently used standard public key encryption methods will not be able to withstand attacks by quantum computers. The transition to quantum computer resistant methods has to be implemented well before the actual availability of quantum computers. Otherwise, today's communication can simply be stored and be decrypted once quantum computers are available. Therefore, the National Security Agency (NSA) of the United States advises partners and vendors to be prepared for a transition to quantum computer resistant cryptography [14]. QKD is provably resistant against quantum computers. Therefore, the market demand is expected to take off in the future.

Fiber-based QKD solutions for metropolitan access networks (MANs) are commercially available and numerous demonstrational and operational systems have been implemented in the field (see Fig. 4). However, for QKD to be attractive to the end user, also wide area networks (WANs) are needed. In contrast to optical telecommunication, fibers are currently not suitable for quantum WANs since their loss increases exponentially with distance. Classical repeaters destroy quantum information and quantum repeaters are not yet available. For free space links, the loss is mainly due to diffraction and hence increases only linearly with distance. Therefore, satellite links can provide the WAN interconnect of MAN quantum networks (see Fig. 4).

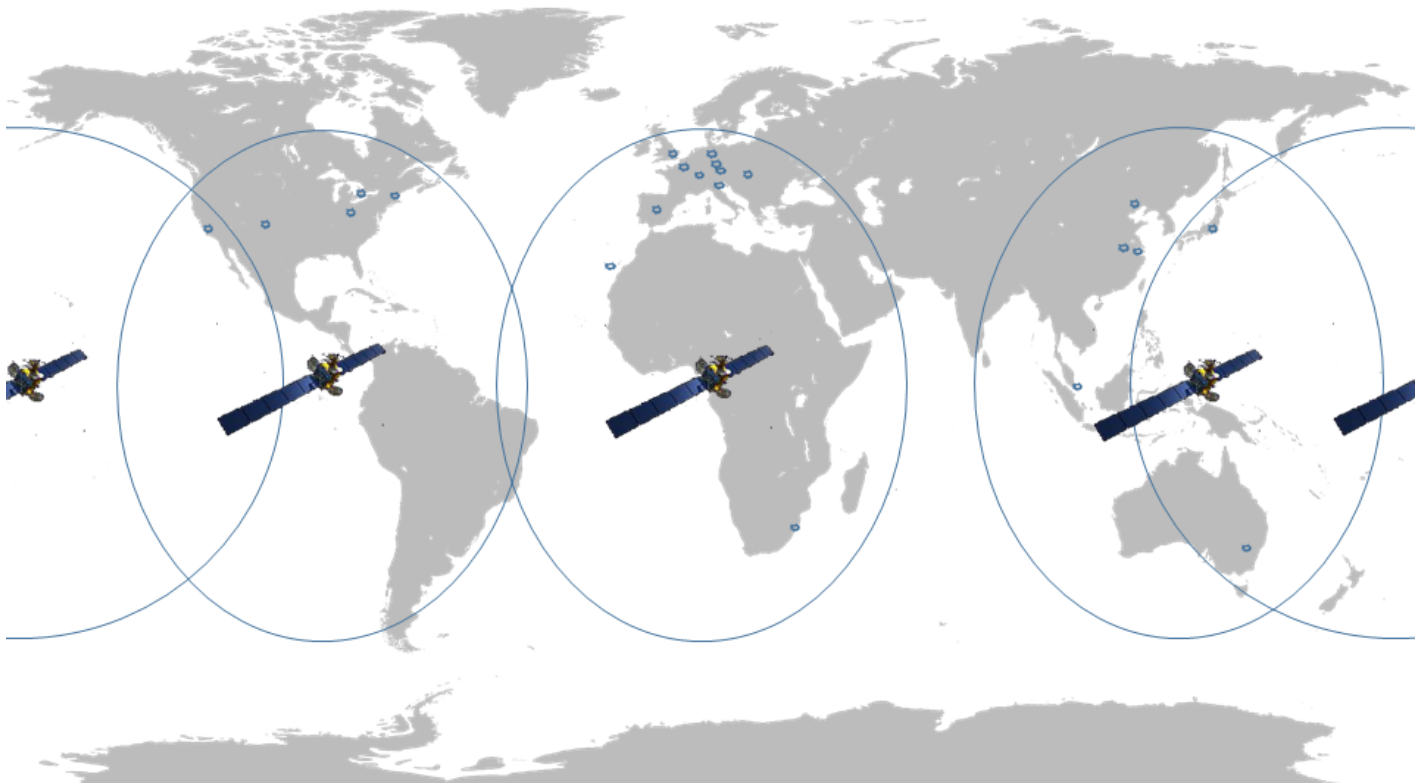

Figure 4: GEO satellites can provide a wide area network (WAN) to interconnect existing quantum metropolitan area networks (MANs).
Blue dots: quantum communication networks and links in the field, blue circles: coverage from geostationary Earth orbit (GEO).

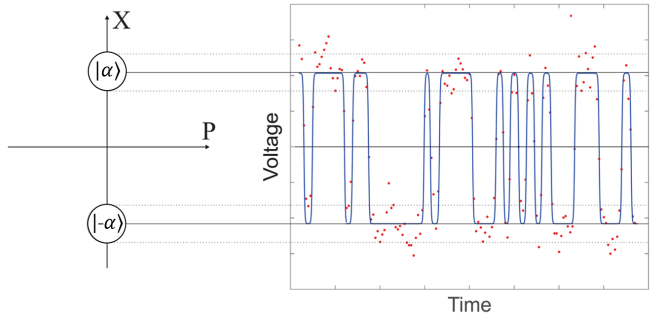

Figure 5: Performance verification of the receiver in the Transportable Adaptive Optics Ground Station. The red dots exemplarily show the quantum-noise limited measurement of two coherent states. The blue curve is the reconstruction signal.

By using existing space technology, a quantum WAN can provide its service for very affordable fees. We have performed a calculation of the service costs assuming realistic prices for hardware, recurring and non-recurring engineering, launch, operational costs and insurance. This gives us user service fees in the order of a few cents per kbit of secure key (see also [15]). Therefore, our proposed quantum WAN is assumed to meet the growing market demand.

## IV. PROJECT STATUS

A feasibility analysis of our approach was successfully concluded in 2012. The validity of our analysis was verified in a terrestrial test bed using quantum communication links across Erlangen in 2013 and 2014 [8]. We have confirmed the suitability of our ground station receiver in a lab environment in 2015 (see Fig. 5). Recently, the commissioning phase of the Transportable Adaptive Optics Ground station (T-AOGS) in Tenerife was successfully concluded including the required functional performance tests [16]. In parallel, first quantum measurements have been performed with the T-AOGS in parallel to the station's commissioning. We are currently analyzing the results of these measurements. According to the outcome of this analysis, test procedures for consecutive measurement campaigns will be defined. This will allow us in the near future to propose specifications for upgrades to LCTs and ground stations. This would allow for optimizing their quantum communication performance while preserving their data communication ability.

## V. CONCLUSION

We have argued for the feasibility of implementing wide area quantum networks at affordable costs using existing hardware for space optical communication. The market of secure communication is assumed to demand for such quantum networks in the future. First investigations of the underlying technology support the applicability of continuous-variable quantum key distribution in optical space communication with homodyne coherent communications. The project will be continued with consecutive measurements of satellite communication links. The aim of the project is to create specifications for future quantum cryptographic payloads as add-ons to already commercially available Laser Communications Systems.

ACKNOWLEDGMENT

We acknowledge Zoran Sodnik for hosting us in the kitchen of the ESA Optical Ground Station (OGS) during our measurement campaign with the Transportable Adaptive Optics Ground Station (T-AOGS) next door.